\documentclass[
 reprint,
 superscriptaddress,
 amsmath,
 amssymb,
 aps,
 prapplied
]{revtex4-2}

\usepackage{graphicx}% Include figure files
\usepackage{dcolumn}% Align table columns on decimal point
\usepackage{bm}% bold math
\usepackage{times}
\usepackage{xcolor}
\usepackage{hyperref}
\hypersetup{hypertex=true, 
	colorlinks=true, 
	linkcolor=blue, 
	anchorcolor=blue, 
	citecolor=blue}
\begin{document}

%\preprint{APS/123-QED}

\title{Broadband Rydberg Atomic Microwave Sensing with 44.6$\,$MHz Instantaneous Bandwidth}% Force line breaks with \\

\author{Yuhan Yan}
\affiliation{%
	State Key Laboratory of Precision Spectroscopy, Institute of Quantum Science and Precision Measurement, School of Physics, East China Normal University, Shanghai, 200062, China
}%
\affiliation{%
	Wangzhijiang Innovation Center for Laser,  Aerospace Laser Technology and System Department, Shanghai Institute of Optics and Fine Mechanics, Chinese Academy of Sciences, Shanghai, 201800, China
}%

\author{Xuejie Li}
\affiliation{%
	State Key Laboratory of Precision Spectroscopy, Institute of Quantum Science and Precision Measurement, School of Physics, East China Normal University, Shanghai, 200062, China
}%
\affiliation{%
	Wangzhijiang Innovation Center for Laser,  Aerospace Laser Technology and System Department, Shanghai Institute of Optics and Fine Mechanics, Chinese Academy of Sciences, Shanghai, 201800, China
}%
\author{Jinyin Wan}
\affiliation{%
	Wangzhijiang Innovation Center for Laser,  Aerospace Laser Technology and System Department, Shanghai Institute of Optics and Fine Mechanics, Chinese Academy of Sciences, Shanghai, 201800, China
}%
\affiliation{%
	Center of Materials Science and Optoelectronics Engineering, University of Chinese Academy of Sciences, Beijing, 100049, China
}%
\author{Xing Xia}
\affiliation{%
	Wangzhijiang Innovation Center for Laser,  Aerospace Laser Technology and System Department, Shanghai Institute of Optics and Fine Mechanics, Chinese Academy of Sciences, Shanghai, 201800, China
}%
\affiliation{%
	Center of Materials Science and Optoelectronics Engineering, University of Chinese Academy of Sciences, Beijing, 100049, China
}%
\author{Haojie Zhao}
\affiliation{%
	Wangzhijiang Innovation Center for Laser,  Aerospace Laser Technology and System Department, Shanghai Institute of Optics and Fine Mechanics, Chinese Academy of Sciences, Shanghai, 201800, China
}%
\affiliation{%
	Center of Materials Science and Optoelectronics Engineering, University of Chinese Academy of Sciences, Beijing, 100049, China
}%
\author{Binghong Yu}
\affiliation{%
	Wangzhijiang Innovation Center for Laser,  Aerospace Laser Technology and System Department, Shanghai Institute of Optics and Fine Mechanics, Chinese Academy of Sciences, Shanghai, 201800, China
}%
\affiliation{%
	Center of Materials Science and Optoelectronics Engineering, University of Chinese Academy of Sciences, Beijing, 100049, China
}%

\author{Jianliao Deng}
\altaffiliation{jldeng@siom.ac.cn}
\affiliation{%
	Wangzhijiang Innovation Center for Laser,  Aerospace Laser Technology and System Department, Shanghai Institute of Optics and Fine Mechanics, Chinese Academy of Sciences, Shanghai, 201800, China
}%
\affiliation{%
	Center of Materials Science and Optoelectronics Engineering, University of Chinese Academy of Sciences, Beijing, 100049, China
}%
\author{Huadong Cheng}
\altaffiliation{chenghd@siom.ac.cn}
\affiliation{%
	Wangzhijiang Innovation Center for Laser,  Aerospace Laser Technology and System Department, Shanghai Institute of Optics and Fine Mechanics, Chinese Academy of Sciences, Shanghai, 201800, China
}%
\affiliation{%
	Center of Materials Science and Optoelectronics Engineering, University of Chinese Academy of Sciences, Beijing, 100049, China
}%
\author{L. Q. Chen}
\altaffiliation{lqchen@phy.ecnu.edu.cn}
\affiliation{%
	State Key Laboratory of Precision Spectroscopy, Institute of Quantum Science and Precision Measurement, School of Physics, East China Normal University, Shanghai, 200062, China
}%
\affiliation{%
	Hefei National Laboratory, Hefei, 230088, China
}%


\raggedbottom
\begin{abstract}
Rydberg atoms have become a promising novel type of microwave sensor due to their excellent physical properties -- broad frequency coverage and large electric dipole moments. High sensitivity and broad instantaneous bandwidth are two indispensable requirements for deployable Rydberg microwave sensors. However, enabling broadband operation while retaining high sensitivity has been a longstanding barrier limiting their applications. We propose and experimentally demonstrate a Rydberg microwave sensor whose instantaneous bandwidth is significantly enhanced via an auxiliary microwave field. By finely modulating the Rydberg energy levels with this field, we broaden the bandwidth substantially while retaining the sensor's inherent high sensitivity. An instantaneous bandwidth of 44.6$\,$MHz ($\pm$22.3$\,$MHz) with a sensitivity of 225.7$\,$nV$\,$cm$^{-1}\,$Hz$^{-1/2}$ is realized in a thermal \(^{87}\)Rb vapor with the local microwave frequency of 16.03$\,$GHz.
Our work delivers concurrent broad instantaneous bandwidth and high sensitivity for Rydberg microwave sensors, paving a technically viable path for their practical deployment in broadband microwave metrology, radar, and wireless communication.
\end{abstract}

\maketitle
\section{INTRODUCTION}
Microwave (MW) detection holds significant importance across multiple domains including scientific research, telecommunications, and radar systems \cite{skolnik2002role,pan2020microwave,sobol1984microwave,tsui2002digital}. Rydberg-atom-based microwave electrometry has emerged as a transformative next-generation sensing technology featuring intrinsic SI traceability, ultra-wide frequency coverage, and antenna-free calibration \cite{saffman2010quantum, fan2015atom, yuan2023quantum, schlossberger2024rydberg, holloway2014broadband, holloway2017atom, holloway2017electric, sedlacek2013atom, anderson2014two, miller2016radio, anderson2016optical, berweger2023closed, cloutman2024polarization}. To facilitate the practical implementation of Rydberg atoms in MW metrology, the superheterodyne detection is proposed \cite{jing2020atomic}, and the two key parameters of the Rydberg MW sensors are typically emphasized: sensitivity and instantaneous bandwidth (IB). Recent studies have predominantly focused on the optimization of these two critical performance metrics \cite{anderson2020rydberg,meyer2020assessment,meyer2021waveguide,prajapati2021enhancement,zhang2022rydberg,liu2022continuous,ouyang2023continuous}. 
However, it remains challenging to achieve simultaneous enhancement of these two core parameters.
The IB of the superheterodyne system is closely related to the full width at half maximum (FWHM) of the electromagnetically induced transparency (EIT) signal. Increasing the Rabi frequencies of the Gaussian beams can indeed broaden the EIT spectrum, this is typically achieved by reducing the beam waist since the available laser power is a limiting constraint in practical systems. This reduction in beam waist consequently diminishes the volume of the light-atom interaction region, decreasing the number of Rydberg atoms and resulting in the degradation of the sensitivity. As a result, preserving high sensitivity while realizing broad IB is a long-standing challenge.

In recent years, there has been extensive research focused on enhancing IB. By employing a dual-beam excitation scheme in a thermal cesium vapor, the IB was increased from 150$\,$kHz to 6.8$\,$MHz. However, this improvement was accompanied by a reduction in sensitivity, which was decreased from 55$\,$nV$\,$cm$^{-1}\,$Hz$^{-1/2}$ to 468$\,$nV$\,$cm$^{-1}\,$Hz$^{-1/2}$ \cite{hu2023improvement}. Through frequency-comb probing, the NIST team demonstrated a 12$\,$MHz IB with 1.91$\,\mu$V$\,$cm$^{-1}\,$Hz$^{-1/2}$ sensitivity \cite{dixon2023rydberg,artusio2024increased}. The extremely high sensitivity of 10$\,$nV$\,$cm$^{-1}\,$Hz$^{-1/2}$ has been demonstrated in a magneto-optical trap (MOT) cold-atom platform,  yet the IB is merely $\pm$2.3$\,$MHz \cite{tu2024approaching}.  It is noteworthy that the study in Ref. \cite{yang2024highly} revealed that the superheterodyne signal originates from two simultaneously six-wave mixing processes, and an IB of $\pm$10.2$\,$MHz was achieved while maintaining a sensitivity of 62$\,$nV$\,$cm$^{-1}\,$Hz$^{-1/2}$. 
In principle, enhancing the IB calls for a higher coupling Rabi frequency. Yet, as demonstrated in Ref. \cite{yang2024highly}, excessively raising coupling Rabi frequency leads to a sharp response dip that hinders continued bandwidth expansion.
Therefore, achieving broader IB while preserving high sensitivity has remained a long-standing challenge, with no major breakthroughs reported in the past two years. This critical impasse calls for innovative approaches to overcome the performance limitation that has hindered the practical deployment of Rydberg MW sensors.

In this paper, we theoretically propose a novel IB enhancement strategy for Rydberg MW sensors and provide the comprehensive experimental validation. By introducing an auxiliary MW field and adjusting its Rabi frequency and frequency detuning, a marked improvement in IB is realized  while preserving high sensitivity. Our experimental results show that we realized an IB of 44.6$\,$MHz ($\pm$22.3$\,$MHz) while achieving an electric field sensitivity of 225.7$\,$nV$\,$cm$^{-1}\,$Hz$^{-1/2}$. This scheme enables substantial IB enhancement while preserving high sensitivity.
\begin{figure}[t!]
	\centering
	\includegraphics[width=1\linewidth]{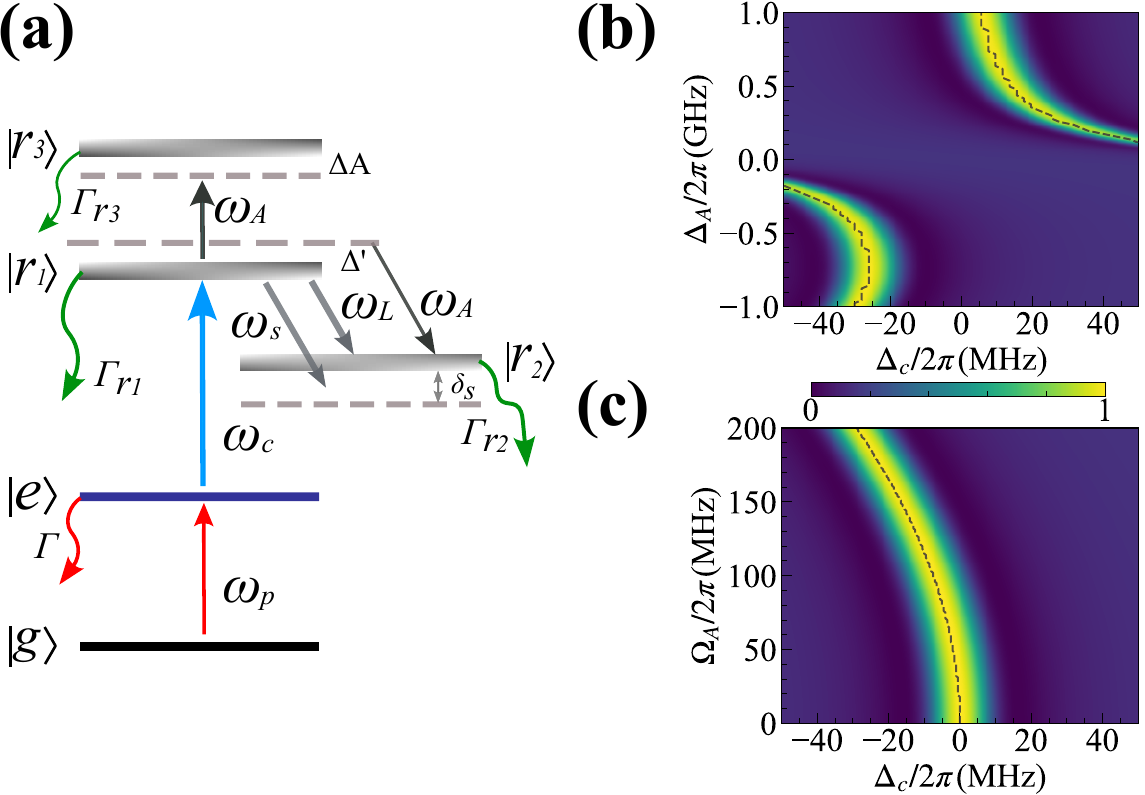}
	\caption{
		(a) Energy level scheme. $\omega_{i}$: $i=p, c, L, s, A$ represent the angular frequency of the probe, coupling, local MW, signal MW, and auxiliary MW fields, respectively. $\Gamma$: The spontaneous decay rate of energy level $|e\rangle$. $\Gamma_{ri}$: $i=1,2,3$ represent the decay rate of the Rydberg states $|r_1\rangle$, $|r_2\rangle$, and $|r_3\rangle$, respectively. $|g\rangle$: $|5S_{1/2}, \ F=2\rangle$, $|e\rangle$: $|5P_{3/2}, \ F'=3\rangle$, $|r_1\rangle$: $|51D_{5/2}, m_J=1/2\rangle$, $|r_2\rangle$: $|52P_{3/2}, m_J=1/2\rangle$, $|r_3\rangle$: $|50F_{7/2}, m_J=1/2\rangle$. The EIT transmission spectrum as a function of $\Delta_A$ when $\Omega_A/2\pi=100\,$MHz and as a function of $\Omega_A$ when $\Delta_A/2\pi=-0.5\,$GHz are shown in (b) and (c), respectively. The color bar indicates the normalized transmission intensity of the probe laser, with brighter yellow representing stronger transmission. The black dashed curve denotes the position of the EIT transmission peak. The EIT transmission spectrum are calculated when $\Omega_p/2\pi=1\,$MHz and $\Omega_c/2\pi=10\,$MHz.
	}
	\label{fig1}
\end{figure}
\section{Theoretical analysis}
Our experimental scheme for enhancing IB is based on a five-level energy structure shown in Fig. \ref{fig1}(a). $\left\vert g\right\rangle$ and $\left\vert e\right\rangle$ is coupled by a probe field ($\omega_p$) with a Rabi frequency $\Omega_p$, $\left\vert e\right\rangle$ and $\left\vert r_1\right\rangle$ are coupled by a coupling field ($\omega_c$) with a Rabi frequency  $\Omega_c$. A local MW field ($\omega_L$, Rabi frequency $\Omega_L$) couple the Rydberg states $\left\vert r_1\right\rangle$ and $\left\vert r_2\right\rangle$. The weak periodic signal MW field ($\omega_s$, Rabi frequency $\Omega_s\ll \Omega_L$) is injected into the system with a frequency detuning of $\delta_s=\omega_s-\omega_L$. An auxiliary MW field ($\omega_A$ , Rabi frequency $\Omega_A$) is coupling the Rydberg states $\left\vert r_1\right\rangle$ and $\left\vert r_3\right\rangle$ with a detuning of $\Delta_A$, the detuning relative to $\omega_{r_1 r_2}$ ($\omega_{ij}$ is the resonant frequency of $|i\rangle$ and $|j\rangle$, and $i, j\in \{g, e, r_1, r_2, r_3\}$) is  $\Delta'$. It should be noted that $\omega_{r_1 r_3}-\omega_{r_1 r_2} \approx 1.5\,$GHz and the transition dipole moment between $|r_1\rangle$ and $|r_3\rangle$ is approximately equal to that between $|r_1\rangle$ and $|r_2\rangle$. Therefore, we must account for not only the coupling between Rydberg states $|r_1\rangle$ and $|r_3\rangle$ induced by the auxiliary MW field, but also its coupling between $|r_1\rangle$ and $|r_2\rangle$. When $\Delta'\gg \delta_s$, the fast-oscillating terms $\Omega_Ae^{-i\Delta't}/2+c.c$ in the Hamiltonian is eliminated. Considering the AC Stark shifts induced by the auxiliary MW field on Rydberg energy levels $\left\vert r_1\right\rangle$ and $\left\vert r_2\right\rangle$, the Hamiltonian of the system after the rotating wave approximation (RWA) is
\begin{figure}[t!]
	\centering
	\includegraphics[width=1\linewidth]{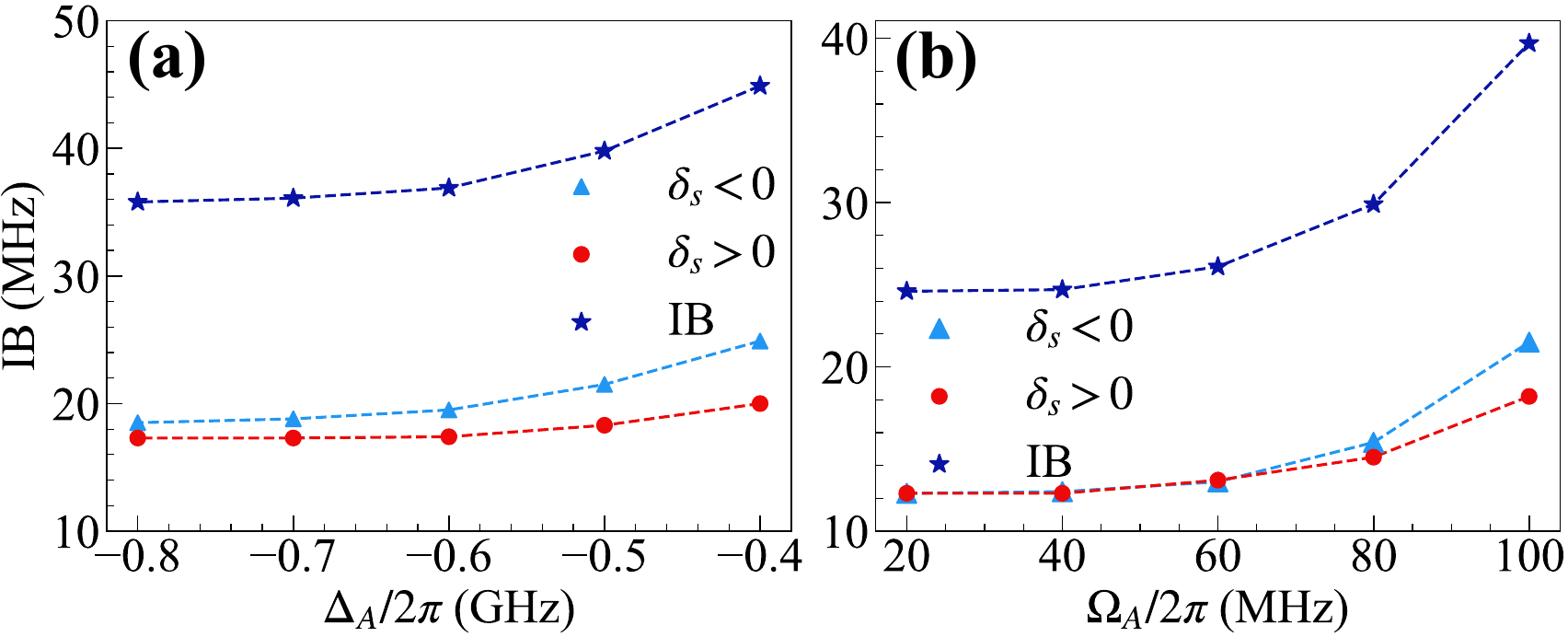}
	\caption{
		The theoretical calculations when $\Omega_p/2\pi=1\,$MHz and $\Omega_c/2\pi=10\,$MHz. The IB as a function of $\Delta_A$ when $\Omega_A/2\pi=100\,$MHz and as a function of $\Omega_A$ when $\Delta_A/2\pi=-0.5\,$GHz are presented in (a) and (b), respectively.
	}
	\label{fig2}
\end{figure}
\begin{equation}
	\begin{aligned}
		\hat{H}_r&=-\Delta_p\hat{\sigma}_{ee}-(\Delta_1+\frac{\Omega_A^2}{4\Delta'})\hat{\sigma}_{r_1r_1}-(\Delta_2-\frac{\Omega_A^2}{4\Delta'})\hat{\sigma}_{r_2r_2}\\
		&-\Delta_3\hat{\sigma}_{r_3r_3}+\left(\right. \frac{\Omega_p}{2}\hat{\sigma}_{ge}+\frac{\Omega_c}{2}\hat{\sigma}_{er_1} +\frac{\Omega_L^*+\Omega_s^* e^{-i\delta_s t}}{2}\hat{\sigma}_{r_1r_2}\\
		&+\frac{\Omega_A}{2}\hat{\sigma}_{r_1r_3}+H.c.\left.\right)
	\end{aligned}
	\label{eqhal}
\end{equation}
where $\Delta_1=\Delta_p+\Delta_c$, $\Delta_2=\Delta_p+\Delta_c-\Delta_L$, $\Delta_3=\Delta_p+\Delta_c+\Delta_A$, and $\Delta_{p,c,L,A}$ are the frequency detuning of the probe, coupling, local MW, and auxiliary MW fields, respectively. $\hat{\sigma}_{i j}=|i \rangle \langle j|$ ($i, j\in \{g, e, r_1, r_2, r_3\} $) is the atomic transition operator which represents the transition from the atomic state $|j \rangle$ to $|i \rangle$.
For such a five-level system, the time evolution of the density matrix is governed by the Lindblad master equation \cite{lindblad1976generators}
\begin{figure*}[t!]
	\centering
	\includegraphics[width=0.8\linewidth]{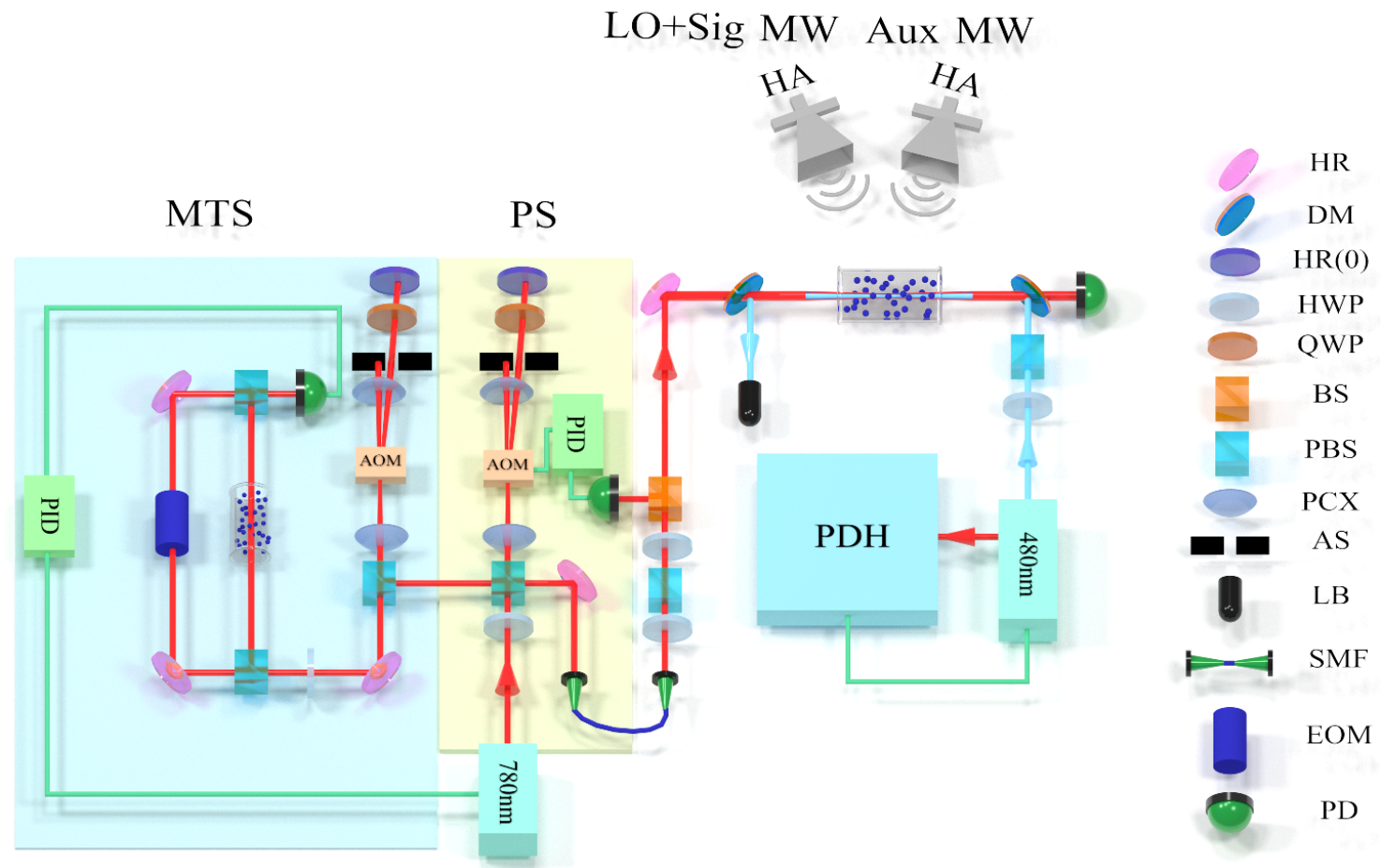}
	\caption{
		Experimental setup. HR: high-reflection mirror; DM: dichroic mirror; HR(0): 0° high-reflection mirror; HWP: half wave plate; QWP: quarter wave plate; BS: beam splitter; PBS: polarization beam splitter; PCX: plano-convex lens; AS: aperture stop; LB: laser block; SMF: single-mode fiber; EOM: electro-optic modulator;
		AOM: acousto-optic modulator; PD: photodetector; PID: proportional-integral-derivative feedback; PDH: Pound-Drever-Hall technique; MTS: modulation transfer spectroscopy; PS: power stabilization; LO: local MW field; Sig: signal MW field; Aux: auxiliary MW field; HA: horn antenna.
	}
	\label{fig3}
\end{figure*}
\begin{equation}
	\begin{aligned}
		\frac{d\rho}{dt}=-i\left[\hat{H}_r,\rho\right]-\frac{1}{2}\left\{\bm{\Gamma}_{spon},\rho\right\}+\bm{\Lambda}
	\end{aligned}
	\label{eqLindblad}
\end{equation}
where $\rho$ is the density matrix of the five-level system, and $\bm{\Gamma}_{spon}$ denotes the spontaneous emission matrix which accounts for the spontaneous decay processes between energy levels, $\bm{\Lambda}$ represents the repopulation term.
Since the probe transmission signal is $\propto\text{Im}(\rho_{ge})$  \cite{Cohen-Tannoudji1996,Boyd2020Nonlinear, PhysRevA.88.033417}, the solution of $\rho_{ge}$ should be obtained to characterize the system response. 
According to the Floquet theory \cite{Eckardt2015,Rodriguez-Vega_2018,rudner2020floquetengineershandbook}, for a system driven by the periodic field $\Omega_s e^{i\delta_st}+c.c$, the density matrix $\rho$
can be decomposed into the following form (saving to the first order)
\begin{equation}
	\rho=\rho^{(0)}+\rho^{(+1)}e^{i\delta_s t}+\rho^{(-1)}e^{-i\delta_s t}
	\label{eqrho}
\end{equation}
Substitute Eq. (\ref{eqhal}) and Eq. (\ref{eqrho}) into Eq. (\ref{eqLindblad}), and utilizing the steady state condition $\partial_t\rho=0$, the analytical solution of $\rho_{ge}^{(0)}$ and $\rho_{ge}^{(\pm 1)}$ can be obtained under the weak probe-field approximation,  which are given by (the detailed derivation provided in the Appendix)
\begin{equation}
	\begin{aligned}
		\rho_{ge}^{(0)} &=\frac{i\Omega_p}{2}\frac{\gamma_{gr_2}\left(\left\vert \Omega_A \right\vert^2+4\gamma_{gr_2} \gamma_{gr_3}\right)+\gamma_{gr_3}\left\vert\Omega_L \right\vert^2}{D(0)} \\
		\rho_{ge}^{(+1)} &=\frac{i\Omega_p}{2}K^*\frac{\gamma_{gr_2}\gamma_{gr_3}\left(\gamma_{gr_3}-i\delta_s\right)}{D(-\delta_s)} \\
		\rho_{ge}^{(-1)} &= \frac{i\Omega_p}{2}K\frac{\gamma_{gr_3} \left(\gamma_{gr_2}-i\delta_s\right)\left(\gamma_{gr_3}+i\delta_s\right)}{D(\delta_s)}
	\end{aligned}
	\label{solrho}
\end{equation}
The definition of $\gamma_{ge}$, $\gamma_{gr_1}$, $\gamma_{gr_2}$, $\gamma_{gr_3}$, $K$, and $D(\delta_s)$ are given in the Appendix. 
Therefore, the superheterodyne response is
\begin{equation}
	\begin{aligned}
		S(\delta_s)=\left\vert \rho_{ge}^{(+1)}-\rho_{ge}^{(-1)*}\right\vert \cos\left(\delta_s t+\varphi\right)
	\end{aligned}
	\label{eq11}
\end{equation} 
where $\varphi=\arg\left[\rho_{ge}^{(+1)}-\rho_{ge}^{(-1)*}\right]$ is the additional phase induced by the signal MW field.  
We define the normalized response amplitude as $S_N(\delta_s)=20\log_{10}\left\vert S(\delta_s)/S(0) \right\vert$, the IB is taken as  the sum of the positive and negative $\delta_s$ where $S_N(\delta_s)$ falls to $-3\,$dB.
For thermal atoms, we should consider the Doppler average, then the atomic coherence is
\begin{equation}
	\begin{aligned}
		\langle \rho_{ge}^{(0,\pm1)}\left(\Delta_p,\Delta_c\right) \rangle=\frac{1}{\sqrt{\pi}u}\int_{-3u}^{3u}\rho_{ge}^{(0,\pm1)}\left(\Delta_p',\Delta_c'\right)e^{-v^2/u^2}dv
	\end{aligned}
	\label{eqthermalInte}
\end{equation}
where $v$ is the velocity of the atomic thermal motion, $u=\sqrt{2k_{B}T/m}$ is the most probability speed, $k_B$ is the Boltzmann constant, $m$ is the mass of single atom, $T$ is the temperature, and
$\Delta_{p,c}'=\Delta_{p,c}-\bm{\vec{k}}_{p,c}\cdot\bm{\vec{v}}$,
$\bm{\vec{k}}_{p,c}$ is the wave vector of the probe and coupling fields, respectively.

We present in Figs. \ref{fig1}(b) and \ref{fig1}(c), respectively, the dependence of the EIT transmission spectrum on the auxiliary MW field frequency detuning $\Delta_A$ and Rabi frequency $\Omega_A$, which can characterize the AC Stark shift of the Rydberg state $|r_1\rangle$. This AC Stark shift modifies the resonance frequencies of Rydberg states $|r_1\rangle$ and $|r_2\rangle$, which in turn alters the coupling between the MW fields (including the local MW and signal MW fields) and the two Rydberg states.
We have theoretically investigate the dependence of the system's IB on the auxiliary MW field detuning $\Delta_A$ and Rabi frequency $\Omega_A$, with the corresponding results are shown in Fig. \ref{fig2}(b, c). These results clearly demonstrate that over a reasonable parameter range, systematic optimization of the auxiliary MW field parameters--specifically reducing $\Delta_A$ or increasing $\Omega_A$--leading to a substantial improvement of IB.  Nevertheless, when the auxiliary MW frequency approaches resonance ($\Delta_A\rightarrow 0$), a significant fraction of Rydberg atoms in $|r_1\rangle$ will be populated into $|r_3\rangle$, reducing the number of Rydberg atoms available to interact with the signal MW field and thus degrading the system sensitivity. For large values of the Rabi frequency $\Omega_A$, the induced AC Stark shifts of Rydberg states $|r_1\rangle$ and $|r_2\rangle$ become substantial, which further detunes the local and signal MW fields from resonance and also reduces the system sensitivity. 
Therefore, in practical measurements, we need to optimize the parameters of the auxiliary MW field to achieve IB enhancement without significantly degrading the sensitivity. Subsequently, we will experimentally demonstrate that using this scheme, we can achieve an IB enhancement while maintaining the sensitivity at the hundred-$\,$nV$\,$cm$^{-1}\,$Hz$^{-1/2}$ level.
\section{Experimental setup}
The experimental setup is shown in Fig. \ref{fig3}. 
The $^{87}$Rb atomic vapor at a room temperature of 21$\,$℃ are excited from a ground state $|g\rangle$ to a Rydberg state $|r_1\rangle$ by two-photon cascade excitation scheme. The 780$\,$nm probe laser  which is continuously driving the atomic transition $|g\rangle \rightarrow |e \rangle  $, is stabilized by the modulation transfer spectroscopy (MTS) . The linewidth of the probe laser is narrowed to 10$\,$kHz after the frequency stabilization. The 480$\,$ coupling laser is locked to a high-finesse Fabry-P\'{e}rot cavity (with a free spectral range of 1.5$\,$GHz and a finesse of 10000) using the Pound-Drever-Hall (PDH) technique, continuously exciting the atoms from $|e\rangle$ to $|r_1\rangle$. We implement power stabilization of the probe beam using an acousto-optic modulator (AOM) combined with a proportional-integral-derivative (PID) feedback system, which effectively suppresses intensity fluctuations.
The probe beam and coupling beam are propagating in opposite directions to reduce the residual Doppler effect. The probe beam is set to a power of 1.7$\,\mu\mathrm{W}$ with a $1/e^2$ radius of 50$\,\mu\mathrm{m}$ at the center of the atomic cell, while the coupling beam with a power of 516$\,\mathrm{mW}$ and a  $1/e^2$ waist radius of 92$\,\mu\mathrm{m}$ at the center of the atomic cell. These yielding the peak Rabi frequecies of the probe and coupling fields are $\Omega_p/2\pi=17.92\,$MHz and $\Omega_c/2\pi=26.26\,$MHz, respectively. The local MW field and signal MW field coupling the Rydberg states $|r_1\rangle$ and $|r_2\rangle$ are combined using a power combiner and emitted from the same horn antenna to the atoms. An auxiliary MW field is emitted to the atoms from another horn antenna. All the fields, including the optical and MW fields, are vertical polarized. To ensure the uniformity  of  MW, the horn antennas are positioned at a distance exceeding 50$\,$cm from the atomic vapor.  Furthermore,  the 5$\,$cm cell is shielded by the microwave anechoic absorbers to reduce the reflections from conductors which might otherwise influence the measurement outcomes. The EIT and Autler-Townes splitting (ATs) signals are detected by an APD (Thorlabs APD130A) with a 3-dB detection bandwidth of 50$\,$MHz. The superheterodyne signal is detected by the same APD and finally acquired by a spectrum analyzer. All the signal generators and the spectrum analyzer are synchronized to a 10$\,$MHz reference signal derived from a high-stability crystal oscillator.
\begin{figure}[t!]
	\centering
	\includegraphics[width=1\linewidth]{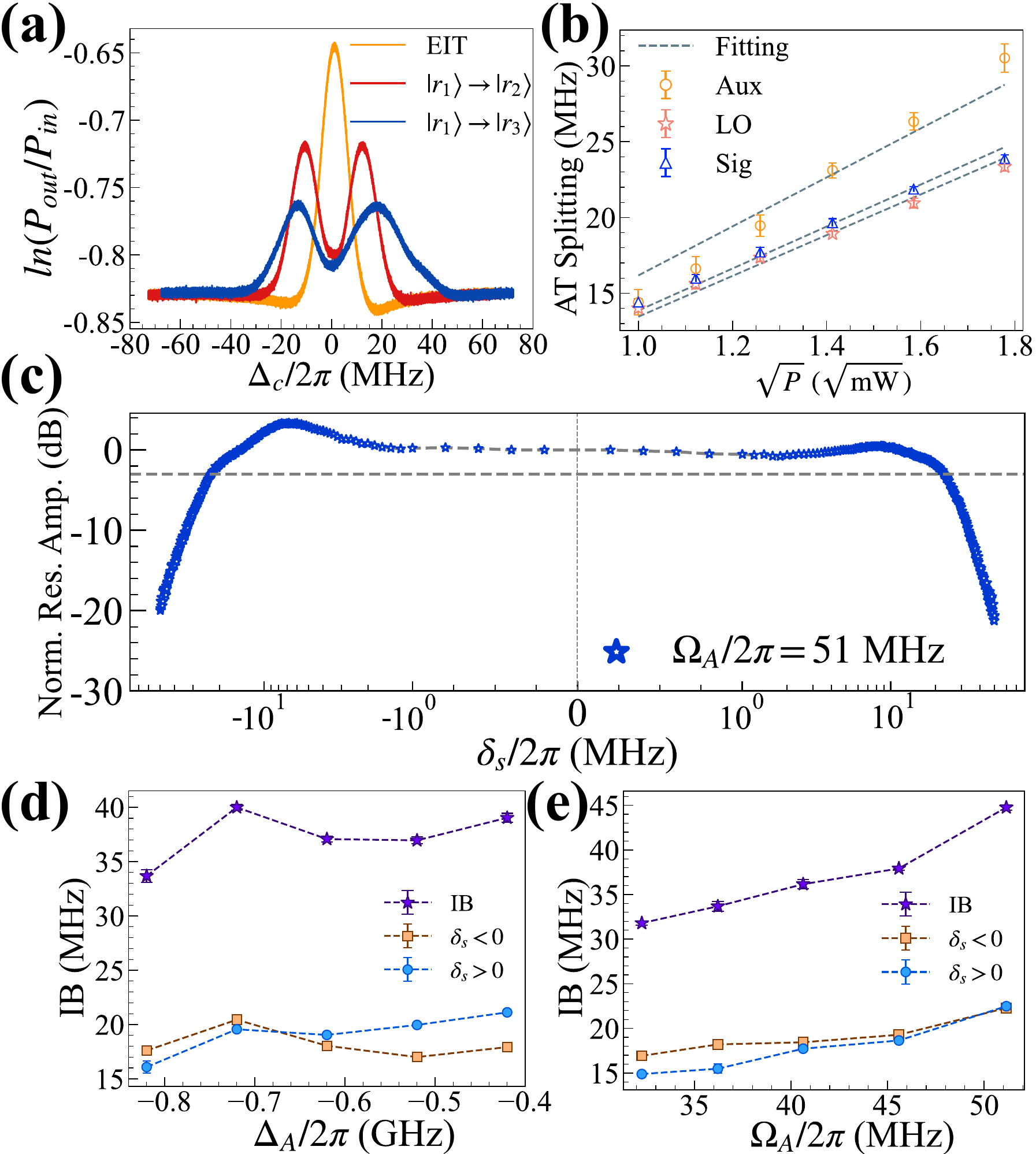}
	\caption{
		(a) The EIT-AT transmission signals. (b) The linear fitting of the AT splitting and the MW field power, including the local, signal, and auxiliary MW fields. (c) The experimental measured normalized response curves when  $\Omega_A/2\pi=51\,$MHz ($\Delta_A/2\pi=-0.72\,$GHz). The gray dashed line represents the 3$\,$dB attenuation of the response amplitude relative to that of $|\delta_s|/2\pi=200\,$kHz. (d) The experimental IB data with different $\Delta_A$ when $\Omega_A/2\pi=51\,$MHz. (e) The experimental IB data with different $\Omega_A$ when $\Delta_A/2\pi=-0.72\,$GHz. The IB is measured when the signal MW power is kept at --45$\,$dBm.
	}
	\label{fig4}
\end{figure}

\section{Experimental results}
In our experiment, we characterize the Rabi frequencies of both the local and auxiliary MW fields and calibrate the signal MW field. The experimentally measured EIT and ATs signals are presented in Fig. \ref{fig4}(a). From the two symmetric transmission peaks of the AT splitting signal, we calibrate the resonance frequencies of the local and auxiliary MW fields with the Rydberg energy levels to be 16.03$\,$GHz and 17.53$\,$GHz, respectively. The linear dependence of the AT splitting $\Delta f$  on the MW power is shown in Fig. \ref{fig4}(b), which satisfy $\Delta f=\Omega_{L,A,s}/2\pi = k_{L,A,s}\sqrt{P_{L,A,s}}$. The slope corresponding to the local, auxiliary, and signal MW fields are $k_{L,A,s}=13.45\,\mathrm{MHz/\sqrt{mW}}$, $k_A=16.17\,\mathrm{MHz/\sqrt{mW}}$, $k_s=13.86\,\mathrm{MHz/\sqrt{mW}}$, respectively. Then, we can determine the corresponding Rabi frequencies of the MW fields at different powers. Subsequently, we measure the frequency response curves of the system with the auxiliary MW field, while maintaining $\Delta_p=\Delta_c=\Delta_L=0$ throughout the measurement process. 
\begin{figure}[t!]
	\centering
	\includegraphics[width=1\linewidth]{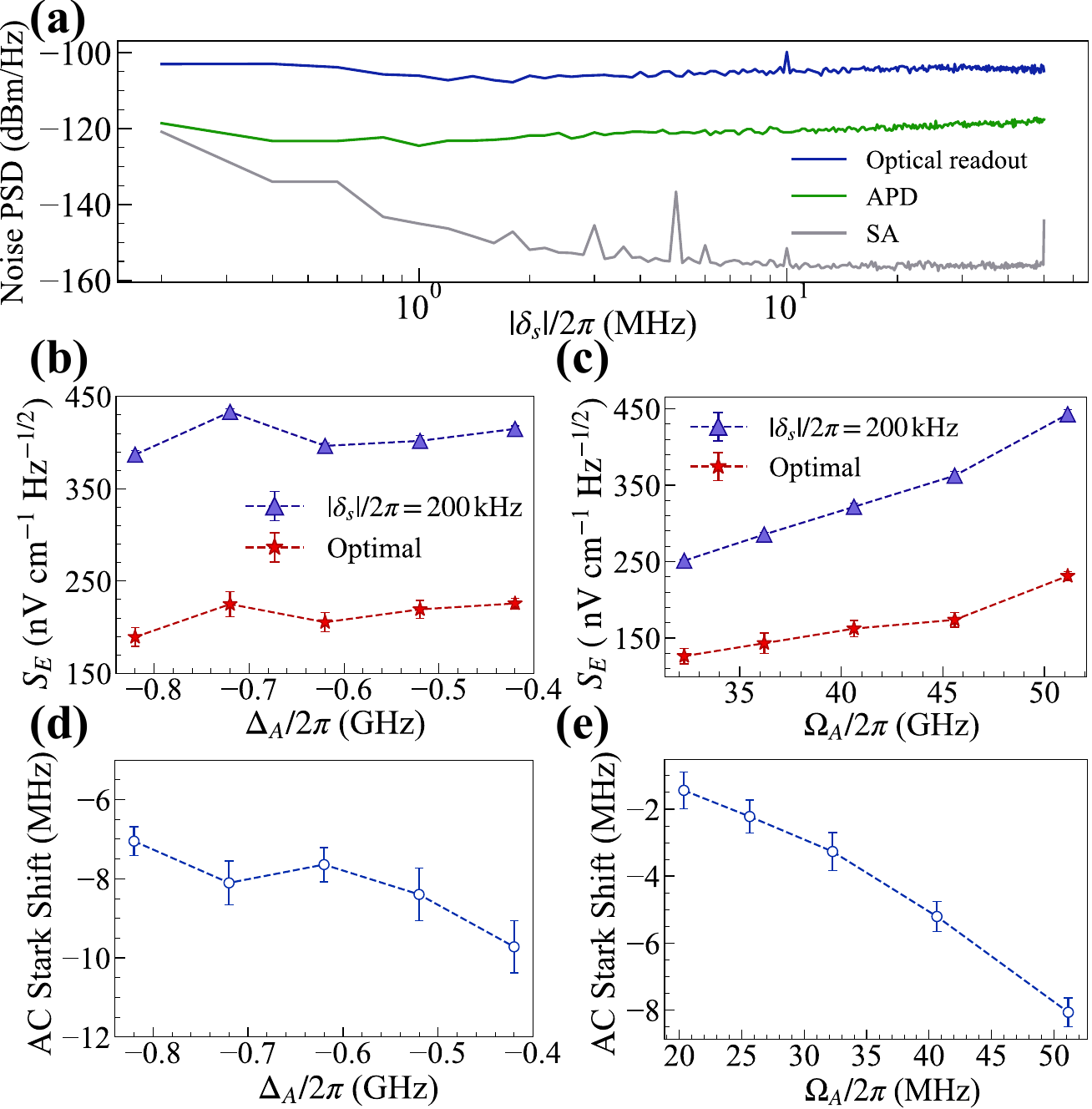}
	\caption{
		(a) The measured noise power spectrum density (PSD) of the Rydberg MW sensor, including the optical readout noise, the noise of the APD, and the noise of the spectrum analyzer (SA). (b) The measured sensitivity data with different $\Delta_A$ when $\Omega_A/2\pi=51\,$MHz. (c) The measured sensitivity data with different $\Omega_A$ when $\Delta_A/2\pi=-0.72\,$GHz.
		(d) The measured AC stark shift of Rydberg state $|r_1 \rangle$ with different $\Delta_A$ when $\Omega_A/2\pi=51\,$MHz. (e) The measured AC stark shift of Rydberg state $|r_1 \rangle$ with different $\Omega_A$ when $\Delta_A/2\pi=-0.72\,$GHz.
	}
	\label{fig5}
\end{figure}
To measure the frequency response, we increase the signal MW field detuning $|\delta_s|$ stepwise while keeping all other parameters constant, and record the response amplitude of the signal at each $\delta_s$ point. We turn on the auxiliary MW field and tune its frequency detuning $\Delta_A$ and Rabi frequency $\Omega_A$ to measure the response curves of the Rydberg MW sensor. 
The response curve for $\Delta_A/2\pi=-0.72\,$GHz and $\Omega_A/2\pi=51\,$MHz is shown in Fig. \ref{fig4}(c). The 3-dB bandwidth reaches 22.3$\,$MHz for both $\delta_s<0$ and $\delta_s>0$, resulting in an IB of 44.6$\,$MHz. We have measured the IB of our system with different  $\Delta_A$ and $\Omega_A$, the results are presented in Fig. \ref{fig4}(d, e). The IB exhibits an increasing trend both when decreasing $\Delta_A$ and when increasing $\Omega_A$, which is in an excellent agreement with the theoretical calculation results shown in Fig. \ref{fig2}(a, b). The IB performance surpasses the state-of-the-art value reported in the literature by more than twofold when the coupling frequency is at resonance in Refs. \cite{yang2024highly} and \cite{11361359}. 

Comprehensive performance assessment of a Rydberg MW sensor requires consideration of more than just IB. The sensitivity remains an equally critical and indispensable performance metric that directly determines the sensor's minimum detectable signal level.
The bandwidth-normalized sensitivity is characterized by the minimum detectable power ($P_{min}$, with the unit of dBm in our experiment) that yields a unity signal-to-noise ratio ($\mathrm{SNR = 1}$) in the optical readout signal, which can be calculated by 
\begin{equation}
	S_E=\eta_s \sqrt{\frac{10^{P_{min}/10}}{f_{RBW}}}
\end{equation}
where $\eta_s=k_sh/\mu_{r_1 r_2}$, $h$ is the Planck constant, $\mu_{r_1  r_2}$ is the transition dipole moment of the Rydberg states $|r_1\rangle$ and $|r_2\rangle$, $f_{RBW}$ is the resolution bandwidth of the spectrum analyzer. According to the experimental results shown in Fig. \ref{fig4}(b), $k_s=13.86\,\mathrm{MHz/\sqrt{mW}}$, then $\eta_s=0.6599\,$V$\,\mathrm{m^{-1}}\,\mathrm{\sqrt{mW}}$. Combined with the noise floor of our system as shown in Fig. \ref{fig5}(a), we can determine the sensitivity at each $\delta_s$ point. 
We present the system sensitivity at $|\delta_s|/2\pi=200\,$kHz and the optimal sensitivity with different $\Delta_A$ and $\Omega_A$ to characterize the sensitivity performance of the Rydberg MW sensor, the measured sensitivity data are presented in Fig. \ref{fig5}(b, c). As illustrated in Fig. \ref{fig5}(b, c), the sensitivity degrades both as the auxiliary MW field approaches resonance and as the Rabi frequency $\Omega_A$ increases.
Our results demonstrate that the optimal sensitivity achieved in our experiment remains over 250$\,$nV$\,$cm$^{-1}\,$Hz$^{-1/2}$.  Specifically, when $\Omega_A/2\pi=51\,$MHz and $\Delta_A/2\pi=-0.72\,$GHz, our system has an optimal sensitivity of 225.7$\,$nV$\,$cm$^{-1}\,$Hz$^{-1/2}$ and the IB of  44.6$\,$MHz ($\pm 22.3\,$MHz).
We have measured the AC Stark shift of Rydberg level $|r_1\rangle$ with different $\Delta_A$ and $\Omega_A$, the results are shown in Fig. \ref{fig5}(d, e), which exhibits the same trend as the sensitivity variation. Therefore, the AC Stark shift of Rydberg states induced by the auxiliary MW field is the factor responsible for the sensitivity degradation in our experiment.

\section{Conclusion}
In conclusion, we theoretically proposed and experimentally demonstrated a novel scheme for enhancing the instantaneous bandwidth of Rydberg microwave sensors while preserving a high level of sensitivity. By introducing an auxiliary microwave field with tuned Rabi frequency and frequency detuning, we achieved a significant enhancement in instantaneous bandwidth with a high sensitivity. Our experimentally demonstrated an instantaneous bandwidth of 44.6$\,$MHz ($\pm22.3\,$MHz) with a sensitivity of 225.7$\,$nV$\,$cm$^{-1}\,$Hz$^{-1/2}$. The instantaneous bandwidth performance has exceeded the best previously reported results by more than twofold. The proposed scheme enables further expansion of the instantaneous bandwidth while striking a favorable balance between broad instantaneous bandwidth and high detection sensitivity, making it highly suitable for microwave detection systems with stringent bandwidth requirements. Our work accelerates the translation of Rydberg microwave sensors from laboratory proof-of-concept to real-world deployment, offering a technically viable route toward broadband microwave metrology, radar detection, and wireless communication applications.
\section{Acknowledgments}
This work is supported by the National
Key Research and Development Program of China
(2024YFA1409404), the National Natural Science Foundation of China (Grants No. 12174409, No. 12304294, No. U23A2075, and No. 12274132), the Fundamental
Research Funds for the Central Universities and Shanghai
Municipal Education Commission (202101070008E00099).

\appendix
\section{The derivation of the superheterodyne response amplitude}
For such a five-level system,  the Hamiltonian under the rotating wave approximation (RWA) $\hat{H}_r$ is 
\begin{equation}
	\begin{aligned}
		\hat{H}_r&=-\Delta_p\hat{\sigma}_{ee}-(\Delta_p+\Delta_c+\frac{\Omega_A^2}{4\Delta'})\hat{\sigma}_{r_1r_1}\\
		&-(\Delta_p+\Delta_c-\Delta_L-\frac{\Omega_A^2}{4\Delta'})\hat{\sigma}_{r_2r_2}\\
		&-(\Delta_p+\Delta_c+\Delta_A)\hat{\sigma}_{r_3r_3}+\left(\right. \frac{\Omega_p}{2}\hat{\sigma}_{ge}+\frac{\Omega_c}{2}\hat{\sigma}_{er_1} \\
		&+\frac{\Omega_L^*+\Omega_s^* e^{-i\delta_s t}}{2}\hat{\sigma}_{r_1r_2}+\frac{\Omega_A}{2}\hat{\sigma}_{r_2r_3}+H.c.\left.\right)
	\end{aligned}
	\label{eqhalApp}
\end{equation}
The terms of $\pm\Omega_A^2/4\Delta'$ is the result for considering the AC Strak shift of the Rydberg states $|r_1\rangle$ and $|r_2\rangle$ induced by the auxiliary MW fied, and the condition of $\Delta'\gg\delta_s$ must be satisfied to eliminate the fast-oscillating terms $\Omega_Ae^{-i\Delta't}/2+c.c$ in the Hamiltonian.
The time evolution of the density matrix is governed by the Lindblad master equation \cite{lindblad1976generators}
\begin{equation}
	\begin{aligned}
		\frac{d\rho}{dt}=-i\left[\hat{H}_r,\rho\right]-\frac{1}{2}\left\{\bm{\Gamma}_{spon},\rho\right\}+\bm{\Lambda}
	\end{aligned}
	\label{eqlindApp}
\end{equation}
$\rho$ is the density matrix of the five-level system, and $\bm{\Gamma}_{spon}$ denotes the spontaneous emission matrix which accounts for the spontaneous decay processes between energy levels, $\bm{\Lambda}$ represents the repopulation term. In our atomic energy level scheme, $\bm{\Gamma}_{spon}$ and $\bm{\Lambda}$ is
\begin{equation}
	\begin{aligned}
		&\bm{\Gamma}_{spon}=\begin{pmatrix}
			\gamma t & 0 &0&0&0 \\
			0& \gamma t+\Gamma&0&0&0 \\
			0 & 0&\gamma t+\Gamma_r&0&0  \\
			0 & 0 &0&\gamma t+\Gamma_r&0  \\
			0&0&0 & 0&\gamma t+\Gamma_r 
		\end{pmatrix}, \\
		&\bm{\Lambda}=\begin{pmatrix}
			\gamma t+\Gamma \rho_{ee}+\Gamma_r\rho_{r_2r_2}&0&0&0&0 \\
			0&\Gamma_r\rho_{r_1r_1}&0&0&0 \\
			0&0&\Gamma_r\rho_{r_3r_3}&0&0 \\
			0&0&0&0&0\\
			0&0&0&0&0
		\end{pmatrix} 
	\end{aligned}
	\label{eqa4}
\end{equation} 
$\gamma t$ is the transit broadening, which depends on the temperature and the light beam waist \cite{MeyerOptimal}. To derive an analytical solution for the superheterodyne signal, we employ the weak-probe field approximation. Thus, to the first order approximation, Eq. (\ref{eqlindApp}) can be rewritten as:
\begin{equation} 
	\begin{aligned}
		\frac{d\bm{X}}{dt}=\bm{FX}+\bm{C}
	\end{aligned}
	\label{eqa3}
\end{equation} 
where
\begin{equation}
	\begin{aligned}
			\bm{X}&=\begin{pmatrix}
				\rho_{ge} \\
				\rho_{gr_1} \\
				\rho_{gr_2} \\
				\rho_{gr_3} \\
			\end{pmatrix}, \bm{F}=
			\begin{pmatrix}
				-\gamma_{ge} & i\Omega_c/2 & 0 & 0 \\
				i\Omega_c/2 & -\gamma_{gr_1} & i\Omega_L/2 & i\Omega_A/2 \\
				0 & i\Omega_L/2 & -\gamma_{gr_2} & 0 \\
				0 & i\Omega_A/2 & 0 & -\gamma_{gr_3}
			\end{pmatrix} \\
			\bm{C}&=\begin{pmatrix}
				i\Omega_p/2 \\
				0 \\
				0 \\
				0 
			\end{pmatrix} 
	\end{aligned}
	\label{eqFXApp}
\end{equation} 
and $\gamma_{ge}=\gamma t+\Gamma/2+i\Delta_p$, $\gamma_{gr_1}=\gamma t+\Gamma_r/2+i(\Delta_p+\Delta_c+\Omega_A^2/4\Delta')$, $\gamma_{gr_2}=\gamma t+\Gamma_r/2+i(\Delta_p+\Delta_c-\Delta_L-\Omega_A^2/4\Delta')$, $\gamma_{gr_3}=\gamma t+\Gamma_r/2+i(\Delta_p+\Delta_c+\Delta_A)$.  According to the Floquet theory \cite{Eckardt2015,Rodriguez-Vega_2018,rudner2020floquetengineershandbook}, for a system driven by a periodic field, the density matrix can be decomposed into the following form (saving to the first order)
\begin{equation}
	\begin{aligned}
		\bm{X}=\bm{X}^{(0)}+\bm{X}^{(+1)}e^{i\delta_s t}+\bm{X}^{(-1)}e^{-i\delta_s t}
	\end{aligned}
	\label{eqXApp}
\end{equation}
Substitute Eq. (\ref{eqhalApp}) and Eq. (\ref{eqXApp}) into Eq. (\ref{eqFXApp}), we have
\begin{equation}
	\begin{cases}
		\begin{aligned}
			&\bm{F}\bm{X}^{(0)}+\bm{C} =0 \\
			&\left(\bm{F}-i\delta_s \bm{I}\right)\bm{X}^{(+1)}+\Omega_s \bm{A}_{+1}\bm{X}^{(0)} =0 \\
			&\left(\bm{F}+i\delta_s \bm{I}\right)\bm{X}^{(-1)}+\Omega_s \bm{A}_{-1}\bm{X}^{(0)} =0 \\
		\end{aligned}
	\end{cases}
	\label{eqFXAPPEq}
\end{equation}
$\bm{I}$ is the identity matrix, and
\begin{equation}
	\bm{A}_{+1}=\begin{pmatrix}
		0 & 0 & 0 & 0 \\
		0 & 0 & i/2 & 0 \\
		0 & 0 & 0 & 0 \\
		0 & 0 & 0 & 0
	\end{pmatrix},
	\bm{A}_{-1}=\begin{pmatrix}
		0 & 0 & 0 & 0 \\
		0 & 0 & 0 & 0 \\
		0 & i/2 & 0 & 0 \\
		0 & 0 & 0 & 0
	\end{pmatrix}
	\label{eqa7}
\end{equation}
By solving Eq. (\ref{eqFXAPPEq}), we obtain the analytical solution for $\rho_{ge}$ as
\begin{equation}
	\begin{aligned}
		\rho_{ge}^{(0)} &=\frac{i\Omega_p}{2}\frac{\gamma_{gr_2}\left(\left\vert \Omega_A \right\vert^2+4\gamma_{gr_2} \gamma_{gr_3}\right)+\gamma_{gr_3}\left\vert\Omega_L \right\vert^2}{D(0)} \\
		\rho_{ge}^{(+1)} &=\frac{i\Omega_p}{2}K^*\frac{\gamma_{gr_2}\gamma_{gr_3}\left(\gamma_{gr_3}-i\delta_s\right)}{D(-\delta_s)} \\
		\rho_{ge}^{(-1)} &= \frac{i\Omega_p}{2}K\frac{\gamma_{gr_3} \left(\gamma_{gr_2}-i\delta_s\right)\left(\gamma_{gr_3}+i\delta_s\right)}{D(\delta_s)}
	\end{aligned}
	\label{solrho}
\end{equation}
$K$ and $D(\delta_s)$ are defined as
\begin{equation}
	\begin{aligned}
		&K=\frac{\left\vert \Omega_c \right\vert^2\Omega_L\Omega_s^*}{D(0)} \\
	&D\left(\delta_s\right) \\
	&=\left(\gamma_{ge}+i\delta_s\right)\left(\gamma_{gr_2}	+i\delta_s\right)\left\vert\Omega_A\right\vert^2\\
	&+\left(\gamma_{gr_3}+i\delta_s\right)\left[\left(\gamma_{ge}+i\delta_s\right)\left\vert\Omega_L\right\vert^2+\left(\gamma_{gr_2}+i\delta_s\right)\left\vert\Omega_c\right\vert^2\right]\\
	&+4\left(\gamma_{ge}+i\delta_s\right)\left(\gamma_{gr_1}+i\delta_s\right)\left(\gamma_{gr_2}+i\delta_s\right)\left(\gamma_{gr_3}+i\delta_s\right)
	\end{aligned}
	\label{eqKApp}
\end{equation}
For thermal atoms, we should consider the Doppler average, then the atomic coherence term $\bm{X}$ is
\begin{equation}
	\begin{aligned}
		\langle \bm{X}\left(\Delta_p,\Delta_c\right) \rangle=\frac{1}{\sqrt{\pi}u}\int_{-3u}^{3u}\bm{X}\left(\Delta_p',\Delta_c'\right)e^{-v^2/u^2}dv
	\end{aligned}
\end{equation}
where $v$ is the velocity of the atomic thermal motion, $u=\sqrt{2k_{B}T/m}$ is the most probability speed, $k_B$ is the Boltzmann constant, $m$ is the mass of single atom, $T$ is the temperature, and 
\begin{equation}
	\begin{aligned}
			\Delta_p'&=\Delta_p-\bm{\vec{k}}_p\cdot\bm{\vec{v}} \\
			\Delta_c'&=\Delta_c+\bm{\vec{k}}_c\cdot\bm{\vec{v}}
	\end{aligned}
\end{equation}
$\bm{\vec{k}}_{p,c}$ is the wave vector of the probe and coupling fields, respectively. Notably, under the strong probe field regime, analytical solutions to the system dynamics are mathematically intractable. Therefore, to accurately derive the frequency response of the superheterodyne signal, we perform numerical simulations by solving the Lindblad master equation.
%\nocite{*}
\twocolumngrid
\bibliography{BandwidthEnhanced}% Produces the bibliography via BibTeX.

\end{document}